\documentclass[%
 reprint,
superscriptaddress,
 amsmath,amssymb,
pra,
]{revtex4-2}

\usepackage[utf8]{inputenc}
\usepackage{listings}
\usepackage[inkscapelatex=false]{svg}
\usepackage{amsmath}
\usepackage{subfigure}
\usepackage{tikz}
\usepackage{algorithm}
\usepackage{algpseudocode}
\usepackage{algorithmicx}
\usepackage{theorem}
\usepackage{physics}
\usepackage{amsmath}
\usepackage{overpic}
\usepackage{tikz}
\usepackage{mathdots}
\usepackage{threeparttable}
\usepackage{yhmath}
\usepackage{cancel}
\usepackage[title]{appendix}
\usepackage{url}
\usepackage{color}
\usepackage{siunitx}
\usepackage{array}
\usepackage{multirow}
\usepackage{amssymb}
\usepackage{gensymb}
\usepackage{tabularx}
\usepackage{extarrows}
\usepackage{booktabs}
\usetikzlibrary{fadings}
\usetikzlibrary{patterns}
\usetikzlibrary{shadows.blur}
\usetikzlibrary{shapes}
\usepackage{graphicx}
\usepackage{dcolumn}
\usepackage{bm}
\usepackage{bbm}
\newtheorem{definition}{Definition}

\begin{document}

\preprint{APS/123-QED}

\title{Scalable Program Implementation and Simulation of the Large-Scale Quantum Algorithm: $1024\times 1024$ Quantum Linear Solver and Beyond}

\author{Zhao-Yun Chen}
 \email{chenzhaoyun@iai.ustc.edu.cn}
\author{Cheng Xue}
 \affiliation{Institute of Artificial Intelligence, Hefei Comprehensive National Science Center, Hefei, Anhui, 230026, P. R. China}
\author{Xi-Ning Zhuang} 
 \affiliation{CAS Key Laboratory of Quantum Information, University of Science and Technology of China, Hefei, Anhui, 230026, P. R. China}  
 \affiliation{Origin Quantum Computing Company Limited, Hefei, Anhui, 230026, P. R. China}
\author{Tai-Ping Sun}
\author{Huan-Yu Liu}
 \affiliation{CAS Key Laboratory of Quantum Information, University of Science and Technology of China, Hefei, Anhui, 230026, P. R. China}
 \affiliation{CAS Center For Excellence in Quantum Information and Quantum Physics, University of Science and Technology of China, Hefei, Anhui, 230026, P. R. China}
  \affiliation{Hefei National Laboratory, Hefei, Anhui, 230088, P. R. China}
\author{Ye Li} 
 \affiliation{Origin Quantum Computing Company Limited, Hefei, Anhui, 230026, P. R. China}
\author{Yu-Chun Wu}
 \email{wuyuchun@ustc.edu.cn}
 \affiliation{Institute of Artificial Intelligence, Hefei Comprehensive National Science Center, Hefei, Anhui, 230026, P. R. China}
 \affiliation{CAS Key Laboratory of Quantum Information, University of Science and Technology of China, Hefei, Anhui, 230026, P. R. China}
 \affiliation{CAS Center For Excellence in Quantum Information and Quantum Physics, University of Science and Technology of China, Hefei, Anhui, 230026, P. R. China}
 \affiliation{Hefei National Laboratory, Hefei, Anhui, 230088, P. R. China}
\author{Guo-Ping Guo}
  \email{gpguo@ustc.edu.cn} 
 \affiliation{Institute of Artificial Intelligence, Hefei Comprehensive National Science Center, Hefei, Anhui, 230026, P. R. China}
 \affiliation{CAS Key Laboratory of Quantum Information, University of Science and Technology of China, Hefei, Anhui, 230026, P. R. China}
 \affiliation{CAS Center For Excellence in Quantum Information and Quantum Physics, University of Science and Technology of China, Hefei, Anhui, 230026, P. R. China}
  \affiliation{Hefei National Laboratory, Hefei, Anhui, 230088, P. R. China}
 \affiliation{Origin Quantum Computing Company Limited, Hefei, Anhui, 230026, P. R. China}



\date{\today}

\begin{abstract}
Program implementation and simulation are essential for research in the field of quantum algorithms. However, complex and large-scale quantum algorithms can pose challenges for existing quantum programming languages and simulators. Here, we present a scalable program implementation of the quantum walk on a sparse matrix and the quantum linear solver based on the quantum walk. Our implementation is based on a practical scenario in which the sparse matrix is stored in the compressed-sparse-column format in quantum random access memory. All necessary modules are implemented unitarily and are ensured to be decomposed at the quantum gate level, including implementing a quantum binary search and a modification of the original algorithm. The program is validated using a highly efficient quantum circuit simulator which is based on the register level and sparse state representation. With only a single core, we simulate the quantum walk on a 16384-dimensional matrix with 582 qubits in 1.1 minutes per step, as well as a quantum linear solver up to 1024 dimensions and 212245 steps in 70 hours. Our work narrows the gap between the simulation of a quantum algorithm and its classical counterparts, where the asymptotic complexity of our quantum linear solver simulation approximates a classical linear solver. These program implementation and simulation techniques have the potential to expand the boundary of numerical research for large-scale quantum algorithms, with implications for the development of error-correction-era quantum computing solutions.
\end{abstract}

\maketitle


\section{\label{sec:introduction}Introduction}

Over the past few decades, there has been significant development in the field of quantum algorithms. One notable example is the quantum walk, which has shown itself to be a powerful tool with a wide range of applications\cite{qw1,qw2,qw3,qw4,qw5}. In particular, the quantum walk on a sparse matrix can be used to implement Hamiltonian simulation and the quantum linear solver\cite{ChildsQLSS}. As a follow-up, the quantum linear solver is often utilized as the fundamental "quantum speedup" subprocess for many other quantum algorithms, including quantum machine learning\cite{ReadFinePrint,QML1,QML2,QML3} and quantum accelerated equation solvers\cite{lloyd_differential_equation,quantum_fem,liujinpeng_nonlinear_differential,XueCheng,qfvm,QPDE}, which are promising to achieve the quantum advantage in real-world tasks.

When these algorithms claim to achieve quantum speedup in terms of asymptotic complexity, it is essential to implement programs and conduct classical simulations to assess their actual performance. However, implementing the quantum walk with existing quantum programming languages and simulators\cite{QPL0,QPL1,QPL2,QPL3,QPL4,QPL5} can be difficult. One major challenge is that the unitary implementation of the quantum walk has not been fully clarified in previous works\cite{QWSim1,ChildsQLSS}, particularly in terms of how to implement all required oracles. Other difficulties include a lack of support for quantum random access memory (QRAM)\cite{QRAM1,QRAM2,QRAM3}, the implementation of quantum arithmetics, management of ancilla resources, and the inability to simulate a large-scale quantum program with more than 100 qubits and a sufficiently large depth. 




In this study, we propose a practical and scalable program implementation method for the quantum walk. To increase practicality, we store the sparse matrix in the QRAM using the compressed-sparse-column (CSC) format\cite{scipy}, which is a common format in classical scientific computing\cite{CSC,CSC2}. The access of the matrix is completely based on the QRAM queries. To extract the sparsity information of the matrix, the quantum computer must find the position of a column in the compressed storage providing the column indices\cite{QSim}. When the column indices are stored in sorted order, we choose the quantum binary search to implement this oracle. We introduce how a while loop of ``quantum data, quantum control''\cite{QIf3} can be performed coherently in a quantum computer\cite{QIf1,QIf2} and the method for automatic management of ancilla registers. Based on the quantum binary search, we propose a modified version of the original quantum walk. This version unitarily implements the sparsity oracle by adding two extra registers, which adapts to the cases where the oracle receives unexpected inputs.

The program is scalable and can adapt to inputs of various sizes, including the dimensions of the matrix, its sparsity, and the size of the matrix elements. We execute the quantum walk program using a quantum circuit simulator on randomly generated band matrices with dimensions ranging from 16 to 16384. These problems require a circuit size that exceeds the limit of all existing simulators, so we propose a novel method called the \textit{sparse state simulator} to simulate the quantum circuit on the register level and only store the nonzero components of the quantum state. The basic operations of this simulator include quantum arithmetic\cite{QArith1,QArith2,QArith3,QArith4}, QRAM queries, conditional rotation, and other essential operations, which can all be simulated as a whole on the register level. The complexity of the simulation is mainly based on the number of nonzero states and is not limited by the number of qubits. On a single core, each quantum walk step uses 1.1 minutes for a $16384\times 16384$ matrix and 1.2 seconds for a $1024\times 1024$ matrix, with a maximum of 582 working qubits. We also program and test Chebyshev's approach of the CKS (Childs, Kothari, Somma) quantum linear solver\cite{ChildsQLSS} on matrices up to 1024 dimensions. As a result, the $1024\times 1024$ linear solver takes 70 hours to simulate on a single core.

All techniques developed in this paper can be a reference for future research on large-scale quantum algorithms, including but not limited to quantum machine learning, and quantum differential equation solvers which may use hundreds of, even thousands of qubits. 

This paper is organized as follows. Section~\ref{sec:preliminaries} overviews our work, and introduces some fundamentals about the quantum walk and the sparse storage in the QRAM. Section~\ref{sec:os} firstly implements a quantum version of the binary search algorithm, then construct a modified version of $O_s$, an input oracle required by the quantum walk algorithm. Section~\ref{sec:qw} scalably implements the remaining parts of the quantum walk. Section~\ref{sec:classical} introduces a novel quantum circuit simulator and shows the numerical results on various sizes of matrices. Later on, we also analyze the reason why we can simulate a quantum algorithm with such a number of qubits. Section~\ref{sec:summary} summarizes the results and outlooks for future works including the scalable implementation of other algorithms, and also the classical simulation on the quantum register level.

\section{Preliminaries\label{sec:preliminaries}}

\begin{figure*}[htbp]
    \centering    
    \includegraphics[width=0.7\linewidth]{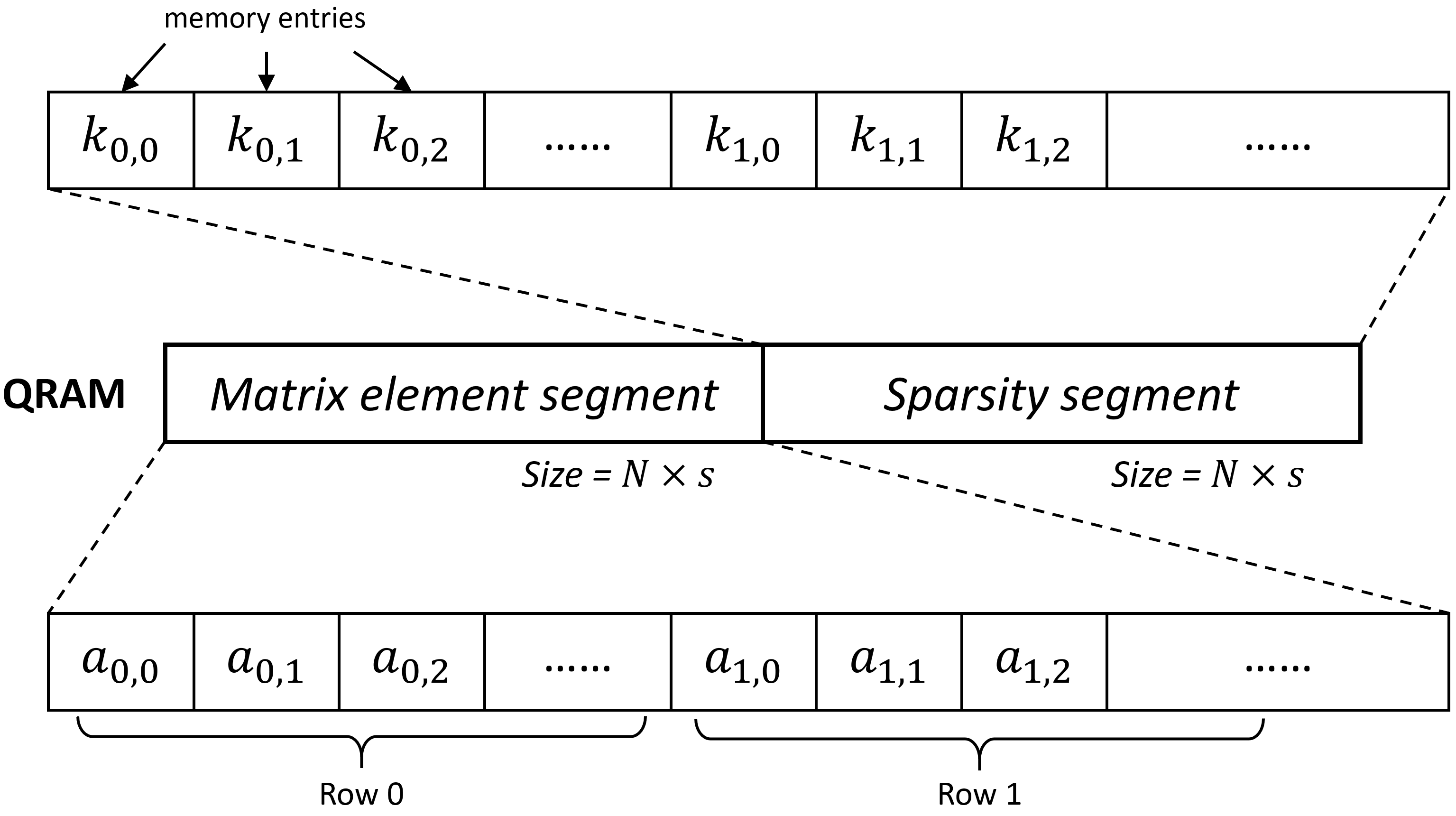}
    \caption{A sparse matrix stored in QRAM in compressed-sparse-column format. It contains two continuous segments: the matrix element segment and the sparsity segments. The matrix element segment compactly stores $s$ elements per row for all $N$ rows. The sparsity segment is the column index of the matrix element corresponding to the same position in the matrix element segment.}
    \label{fig:data structure}
\end{figure*}

\subsection{Overview} 

The quantum walk on a sparse matrix holds great potential for achieving quantum speedup on practical problems such as solving linear systems. If one hopes to apply this algorithm to real-world problems, it is crucial to consider how the input matrix is provided. In this study, we focus on the practical scenario, in which the sparse matrix is stored in the sparse form in the quantum random access memory. This means that all the subprocesses required for a quantum walk, also known as \textit{oracles}, can only be obtained by accessing the QRAM. Therefore, our quantum program implementation is based on QRAM access and arithmetic operations to process data in quantum parallel. With all the subprocesses implemented in a standard manner, we ensure that this program can be reproduced on future quantum computers and that its time complexity is consistent with the original quantum walk theory.

To implement the sparsity oracle by accessing the CSC sparse matrix, we designed a quantum binary search subroutine, which can find the corresponding sparse number according to the column number of a non-zero element. However, the quantum binary search may produce garbage when the input is not included in the search list, which is common in walking multiple steps. Therefore, we also adjusted the form of the quantum walk operator by increasing two extra registers to accept all inputs.

Finally, we show the results of this program executed in a new quantum circuit simulator, including the quantum walk of a 16384-dimensional matrix and the linear system solving of the 1024-dimensional matrix. By comparing with the theoretical quantum walking results, we verify the correctness of the program.




\subsection{Quantum walk on a sparse matrix}

A sparse matrix is a kind of matrix whose zeros occupy most positions, which is very common in various fields. Many works proposed methods to achieve quantum speedup in the solution of sparse linear systems. A formal problem statement was proposed in \cite{ChildsQLSS} and we will follow these definitions throughout this paper. To input the matrix $A$, two oracles are required. One is $O_A$, named as ``matrix element oracle'', which is 
\begin{equation}
    O_A|j\rangle|k\rangle \mapsto |j\rangle|k\rangle|A_{jk}\rangle,
\end{equation}
where $A_{jk}$ is the matrix element in row $j$ and column $k$, and $|A_{jk}|\leq 1$. The other is $O_s$, named as ``sparsity oracle'', which is
\begin{equation}
    O_s|j\rangle|k\rangle \mapsto |j\rangle|l_{j,k}\rangle,
\end{equation}
where the $l_{j,k}$-th nonzero element is at column $j$. It is worth mentioning that $O_s$ is an in-place operation, which means the inverse function (from $l_{j,k}$ to $k$) is also available.

The quantum walk is acted on four quantum registers, that is $\mathcal{H}^N\otimes\mathcal{H}^2\otimes\mathcal{H}^N\otimes\mathcal{H}^2$. The walk operator is a unitary
\begin{equation}
    U=S\left(2 T T^{\dagger}-\mathbbm{1}\right),
\end{equation}
where $T$ is an isometry $T=\sum_{j \in[N]}\left|\psi_{j}\right\rangle \left\langle j\right|$ and $S$ the swap between the first two and the last two registers. And we have 

\begin{eqnarray}
\left|\psi_{j}\right\rangle=|j, 0\rangle &\otimes& \frac{1}{\sqrt{d}} \sum_{k \in[N]: A_{j k} \neq 0} \\
& & \left(\sqrt{A_{j k}^{*}}|k, 0\rangle+\sqrt{1-\left|A_{j k}\right|}|k, 1\rangle\right) \nonumber.
\end{eqnarray}
Repeating $U$ to extract the information in $A$ is the essential process of the quantum walk.

\subsection{Sparse matrix storage model in the quantum random access memory}

A sparse matrix is a matrix where zero occupies most positions. More specifically, a matrix is called $s$-sparse if each column has at most $s$ nonzero elements. There are many formats for storing a sparse matrix, among which the compressed-sparse-column (CSC) format is widely applied in various scenarios and is more intuitive in the quantum walk task. The CSC format compresses each row and only stores the nonzero entries for every row. Meanwhile, the format also stores the column indices of each element correspondingly. 

In Fig.~\ref{fig:data structure}, we demonstrate how a sparse matrix is stored in the QRAM, which includes two continuous segments: the matrix element segment and the sparsity segment. The matrix element segment stores the nonzero data entries $a_{j,l} = A_{jk}$ in each row, and the sparsity segment stores the column indices $k_{j,l}$ of the corresponding data entries. We specify the number of nonzero elements per row as a constant $s$ and fill zeros when the number of nonzeros is less than $s$. To enable the binary search on the sparsity segment, the column indices in the same row are in sorted order.


\begin{figure*}[htbp]
    \centering    
    \includegraphics[width=\linewidth]{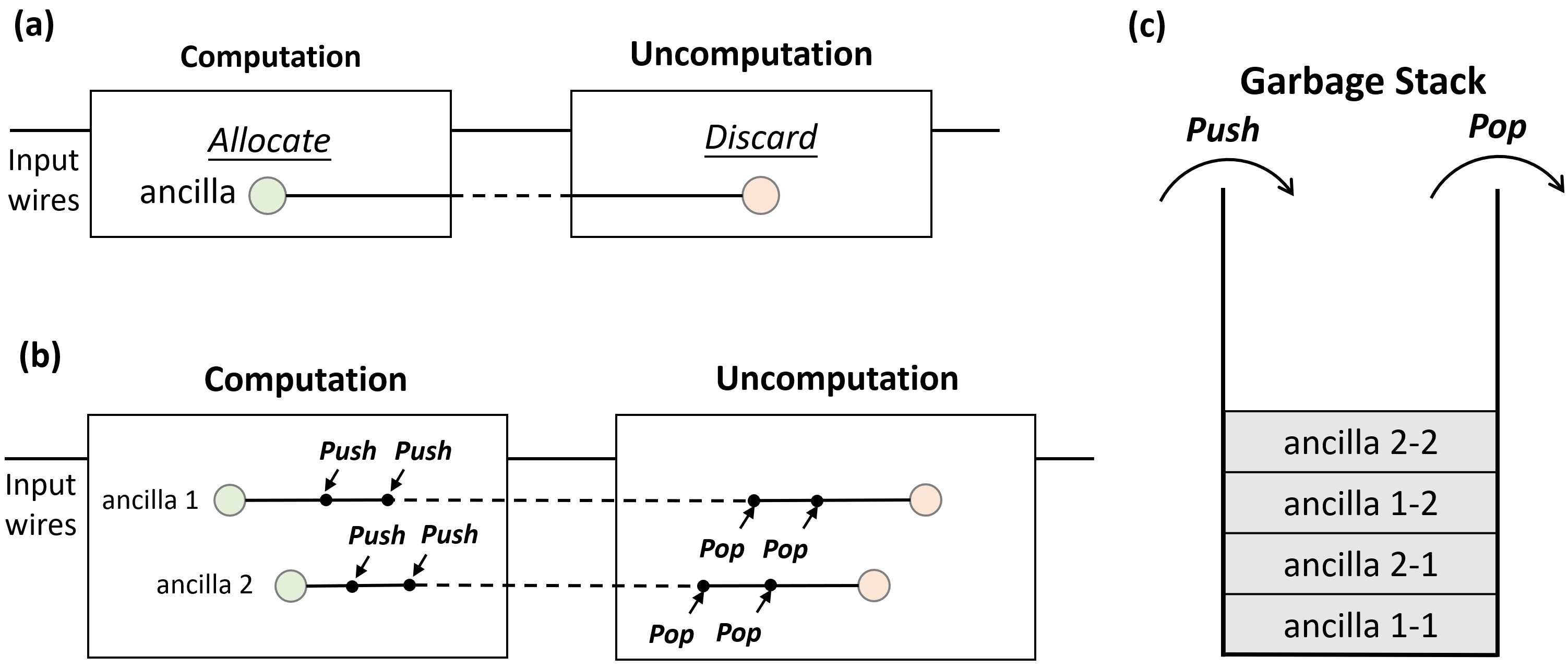}    
    \caption{(a) Computation-uncomputation model, which is classic model of the reversible computing and quantum computing. (b) Push-pop model. In the quantum loop, the garbage created in the loop body is ``pushed'' to the garbage stack when the loop body finishes. It means that this variable is swapped with an empty register at the top of the stack. When applying uncomputation, the variable will be ``poped'' from the stack. While the push operation will maintain the existence the ancilla, it can also be generalised to cases when a temporal variable is overwritten. (c) The garbage stack. This is a stack that logically stores the snapshots of temporal registers that are ``pushed''. As shown in subfigure (b), two snapshots of two ancillae are pushed into the stack. When performing the ``pop`` operations, contents will be returned to the original registers.}
    \label{fig:execution_models}
\end{figure*}

\section{Implementing $O_s$ via the quantum binary search\label{sec:os}}
\subsection{Quantum binary search}
Unitary implementation of $O_s$ is an in-place computation from $|l\rangle$ to $|k_{j,l}\rangle$, which implies that the out-of-place computation $|j,l\rangle\mapsto |j,l\rangle|k_{j,l}\rangle$ and its inverse $|j,k_{j,l}\rangle\mapsto |j,k_{j,l}\rangle|l\rangle$ are both required. The former mapping is straightforward with the CSC format, but the inverse requires the search from $k$ to $l$. \cite{ChildsQLSS} explicitly pointed out that the quantum search is necessary to implement $O_s$, either the quantum binary search in $O(\log s)$ steps when the nonzero column is stored in sorted order, or Grover's search in $O(\sqrt{s})$ steps.

In this section, we show the unitary implementation of the quantum binary search. First, we define the task.

\begin{definition}[Quantum Binary Search]
Given a unique sorted list $d_i$ ($i=0,1,...,N-1$) compactly stored in the QRAM, quantum binary search (QBS) is a unitary such that
\begin{equation}
    \mathrm{QBS}|a\rangle|d\rangle|j\rangle = |a\rangle|d\rangle|j\oplus i(d)\rangle,
\end{equation}
where $|a\rangle$ is the address of $d_0$, $|d\rangle$ the target, $|j\rangle$ any input, $\oplus$ the bitwise XOR, and
$$
i(d)=\left\{\begin{array}{cl}
    0 & \text{if } d \text{ is not in the list} \\
    i & \text{if } d[i] = d
\end{array}\right..
$$
\end{definition}

Quantum binary search is a semi-quantum process, which means that it is a quantum circuit implementation of the classical binary search. However, it is nontrivial to convert the recursive process in the binary search to its quantum version. This is called the quantum recursion of the ``quantum data, quantum control'' model\cite{QIf3}, where the recursion process is determined coherently by a quantum variable. 

Let us start with a simple and classic model of reversible programming\cite{Rev0}, the computation-uncomputation model, shown in Fig.~\ref{fig:execution_models}(a). The \textit{computation} is usually not reversible, therefore some ancillae will be allocated during the computation and be left as the garbage after computation finishes. Later, a reverse \textit{uncomputation} is performed, returning the garbage to the initial state and discarding them safely. This model is also applied to quantum computing languages such as \cite{QPL0, QPL1}.

For the quantum recursion, we extend the computation-uncomputation model and propose the push-pop model which is inspired by the classical reversible programming language\cite{Rev0,Rev1,Rev2,Rev3,Rev4}, shown in Fig.~\ref{fig:execution_models}. The model contains a garbage stack, which has ``push'' and ``pop'' operations. Each push operation will move the immediate value of a variable to the top of the garbage stack. This happens when the state of the ancilla should be overwritten by a new value, such as starting a new loop cycle and all variables are refreshed. The pop operation reverses the push operation, resuming the value in the computation stage. Consequently, garbage stack records the snapshot of variables whenever their value has to be discarded. 

With the push-pop model, now we can implement the quantum binary search, displayed in Alg.~\ref{alg:qbs}. The quantum loop, as shown from step 4 to step 14 in Alg.~\ref{alg:qbs}, runs different steps depending on the input size, thus creating different numbers of ancilla registers. During the quantum loop, some branches have terminated while others may go further. We let the \textit{flag} control the termination of a branch. In line 9, the \textit{flag} register flips when \textit{compareEqual} is true, which represents the loop breaks when the target is found. While terminated branches copy the target to the output register (line 8), other branches have gone to the next loop. In line 13, at end of the loop body, all old variables are pushed to the garbage stack. The original \textit{mid} should be regarded as the temporal garbage used to record the branches route for the uncomputation. In line 15, the loop body reverses, such that all ancilla registers return to the $|0\rangle$ state, and thus can be discarded safely. Finally, we implement the quantum binary search without creating any garbage, and only output registers are changed.

\renewcommand\algorithmicensure{\textbf{Subprocess:}}

\begin{algorithm}[H]
\caption{Implementation of quantum binary search (\textbf{QBS})}
\label{alg:qbs}
\begin{algorithmic}[1]
\Require Offset register \textit{offset}, target register \textit{target}, and output register $j$.
\Require Sparsity constant $s$;
\State Allocate ancilla \textit{left}, \textit{right}, \textit{mid}, \textit{midVal}, \textit{flag}, \textit{compareLess}, and \textit{compareEqual}; 
\State Assign \textit{left} $\gets$ $\rm offset$, \textit{right} $\gets \mathrm{left}+s$;
\State Assign \textit{flag} $\gets$ \textit{true};
\For {i = 0 : $\lceil \log s\rceil$}
    \State (\textit{flag} control) \textit{mid} $\gets$ $(\rm left + right)/2$.
    \State (\textit{flag} control) Apply QRAM query with \textit{mid} and \textit{midVal};
    \State (\textit{flag} control) Compare \textit{midVal} and \textit{target}, and store the result to \textit{compareLess} and \textit{compareEqual};
    \State (\textit{compareEqual} control) Assign output register $j \gets j\oplus \textit{mid}$ ;
    \State (\textit{compareEqual} control) Flip \textit{flag};
    \State (\textit{flag}, \textit{compareLess} control) Swap \textit{mid} and \textit{left};
    \State (\textit{flag} control) Flip \textit{compareLess};
    \State (\textit{flag}, \textit{compareLess} control) Swap \textit{mid} and \textit{right};
    \State Push \textit{mid}, \textit{midVal}, \textit{compareLess}, \textit{compareEqual};
\EndFor
\State Uncompute the above all except step 8.
\State Discard \textit{left}, \textit{right}, \textit{mid}, \textit{midVal}, \textit{flag}, \textit{compareLess}, \textit{compareEqual} safely.
\end{algorithmic}
\end{algorithm}



\subsection{Implementing a modified $O_s$}

\begin{figure*}[ht]
    \centering
    \includegraphics[width=0.8\linewidth]{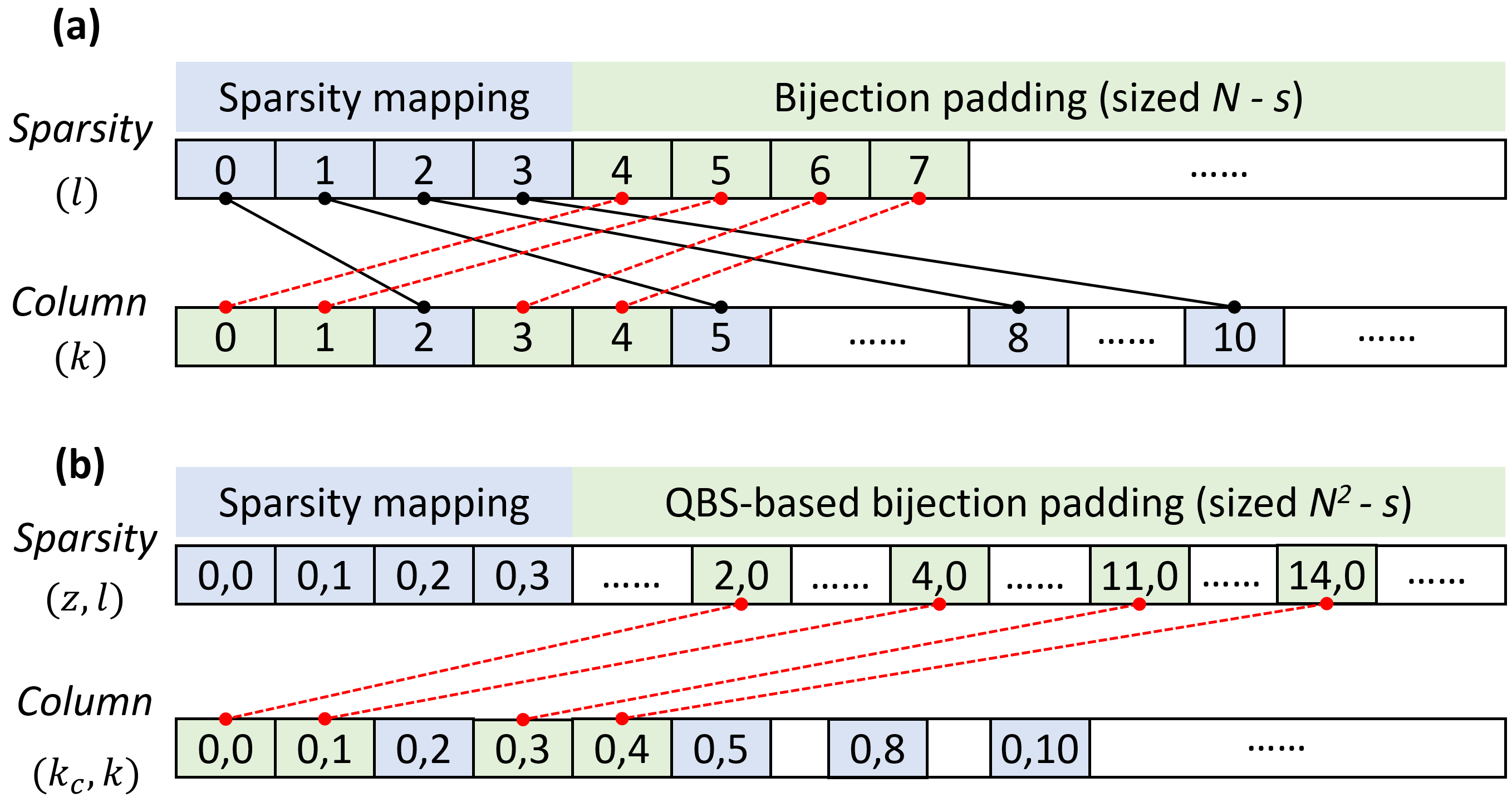}        
    \label{fig:bijection}
    \caption{Two bijection mapping schemes. (a) An intuitive mapping scheme. The input $l$ is divided into two ranges, where $l<s$ represents the common sparsity mapping, and $l\geq s$ is a padding region with $N-s$ elements that maps to the zero columns $k$. (b) The QBS-based mapping scheme. The inputs and outputs are extended to a pair of integers between 0 and $N-1$. The common sparsity mapping region is $(0,l)$ where $l<s$. The padding region is the remaining $N^2-s$ elements of $(z,l)$ that correspond to some certain $(k_c,k)$.}
\end{figure*}


Using the QRAM query, one can implement
\begin{equation}\label{AppendixEqn:QRAM}
    \mathrm{QRAM}|j\rangle|l\rangle|z\rangle = |j\rangle|l\rangle|z\oplus k_{j,l}\rangle,
\end{equation}
where $k_{j,l}$ is the $l$-th column index of row $j$. Using the quantum binary search, one can implement

\begin{equation}\label{AppendixEqn:QBS}
\mathrm{QBS}|j\rangle|k_{j,l}\rangle|z\rangle = |j\rangle|k_{j,l}\rangle|z\oplus \tilde{l}\rangle,
\end{equation}
where $\tilde{l}=l$ if $k_{j,l}$ is a valid column index of row $j$, and $\tilde{l} =0$ if the input $k_{j,l}$ does not exist in row $j$. With Eqn.~(\ref{AppendixEqn:QRAM}) and Eqn.~(\ref{AppendixEqn:QBS}) and some valid input $j$ and $l$, we can implement $O_s$ with the following step:
\begin{equation}\label{AppendixEqn:Process}
\begin{aligned}
    &|j\rangle|l\rangle \\
    (\text{Allocate})\longrightarrow&|j\rangle|l\rangle|0\rangle\\
    (\text{QRAM}){\longrightarrow}&|j\rangle|l\rangle|k_{j,l}\rangle\\
    (\text{QBS}){\longrightarrow}&|j\rangle|0\rangle|k_{j,l}\rangle\\
    (\text{Swap}){\longrightarrow}&|j\rangle|k_{j,l}\rangle|0\rangle\\
    (\text{Deallocate}){\longrightarrow}&|j\rangle|k_{j,l}\rangle.
\end{aligned}
\end{equation}
This process is reversible, indicating that $|j\rangle|k_{j,l}\rangle \rightarrow |j\rangle|l\rangle$ is accepted for valid input $j$ and $k_{j,l}$.

Despite the valid inputs, one should notice that there are also some invalid inputs in the computing process. For example, $k_{j,l}$ may not correspond to any $l$. Also, there may have $l\geq s$, where $s$ is the maximum number of nonzero entries in a row. This will cause the ancilla register introduced in the first step of Eqn.~(\ref{AppendixEqn:Process}) cannot be disentangled in general. If $O_s$ is implemented unitarily and in place, one has to create a bijection mapping that any input $0\leq l<N$ must correspond to a unique $0\leq k<N$. An intuitive example is shown in Fig.~\ref{fig:bijection}(a), demonstrating a matrix with $s=4$, and the nonzero column is located in 2, 5, 8, and 10. These valid inputs are marked as blue blocks, where the invalid inputs are marked as green. The red arrows represent the mapping of $l\geq s$, which map input $l=4$ to $k=0$, $l=5$ to $k=1$, $l=6$ to $k=3$, and so forth. Here, the mapping target for $l\geq s$ is continuously inserted into the space excluded the indices for nonzero columns. 

However, we show that the unitary implementation of this intuitive mapping consumes at most $O(s)$ extra time and space, see Appendix Section~\ref{app:intuitive}. To optimize, we propose a method that only consumes $O(1)$ extra space, and $O(\log s)$ time to implement the same mapping as required by the quantum walk. Our method is to keep the last register as a working register, and modify the original $O_s$ to a 3-register version $O_s'$ with the following steps:
\begin{equation}\label{AppendixEqn:Newprocess}
\begin{aligned}
    &|j\rangle|l\rangle|z\rangle \\
    (\text{QRAM}){\longrightarrow}&|j\rangle|l\rangle|z\oplus k_{j,l}\rangle\\
    (\text{QBS}){\longrightarrow}&|j\rangle|l\oplus y(z\oplus k_{j,l})\rangle|z\oplus k_{j,l}\rangle\\
    (\text{Swap}){\longrightarrow}&|j\rangle|z\oplus k_{j,l}\rangle|l\oplus i(z\oplus k_{j,l})\rangle,
\end{aligned}
\end{equation}
where $i(x)$ denotes the search result produced by the quantum binary search. When $z=0$ and $l<s$, this process is identical to the original $O_s$. The schematic diagram for this mapping is shown in Fig.~\ref{fig:bijection}(b). Alike the above example, we mark the valid inputs as blue blocks and invalid as green, and the mapping is marked as the red arrow. For the first arrow, when $O_s'^\dagger$ takes $k=0$ and $z=0$ input, the QBS cannot find a result and will output $l=z=0$. Then the QRAM process will read from address 0 and outputs column number 2. The final output will be $l=2$ and $k=0$. 

Because the whole process is always closed in these three registers, this bijection mapping is naturally implemented. Instead of mapping from $\mathbb{R}^{N}$ to $\mathbb{R}^{N}$, this method implements the mapping from $\mathbb{R}^{N^2}$ to $\mathbb{R}^{N^2}$. The extra space is the ancilla register used in the QBS, which is an ignorable overhead. Because no extra search is needed, the time cost is mainly due to the QBS, which is $O(\log s)$.

Next, we show how to use this modified $O_s'$ in the implementation of the quantum walk. Based on $O_s'$, we add two $n$-qubit registers $j_c$ and $k_c$, expanding the row and column register $j$ and $k$ correspondingly. Now we rewrite $O_s$ as a mapping between $|j\rangle|l\rangle|z\rangle$ and $|j\rangle|k\rangle|k_c\rangle$. 

Expanding the original matrix $A\in \mathbb{R}^{N\times N}$ to $A'\in \mathbb{R}^{N^2\times N^2}$, where $A'$ has the following property: in row $\overline{j_cj}$, the index of $\overline{zl}$-th column is $\overline{k_ck}$, where $\overline{xy}$ means $Nx+y$. For the expanded matrix and this mapping, column index and its position index are a bijection, which means that it is reversible implementation of $O_s'$ in $A'$. When $j_c=0$, $l<s$ and $z=0$, we can write its output as $k=k_{j,l}$ and $k_c=0$. Therefore, $A$ is exactly located in the left-upper corner of the matrix $A'$, which forms a block encoding of $A$
$$
A'=\begin{pmatrix}
A & *\\
* & *
\end{pmatrix}.
$$
As a consequence, the modified quantum walk of $A$ is the walk on the matrix $A'$.

To be simple enough, we let the original $O_A'|j\rangle|l\rangle$ controlled by the two newly-added registers' with $|j_c\rangle=|k_c\rangle=|0\rangle$. Hence, any $A'_{\overline{j_cj},\overline{k_ck}}=0$ except for $j_c=k_c=0$. This simplifies the structure of $A'$ to 
$$
A'=\begin{pmatrix}
A & 0\\
0 & 0
\end{pmatrix}.
$$
And it is straightforward to see that any walk operators acting on $A'$ is equivalent to a block operation on $A$.

\begin{figure*}[t]
    \centering
    \includegraphics[width=0.7\linewidth]{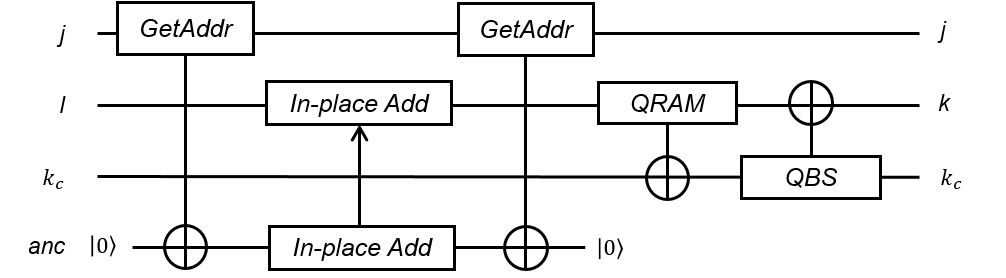}
    \caption{The quantum circuit of $O_s'$.}
    \label{fig:os}
\end{figure*}

Finally, we show the quantum circuit for implementing $O_s'$ in Fig.~\ref{fig:os}.


\section{A scalable implementation of the quantum walk\label{sec:qw}}

\paragraph{Implementation of $\tilde{T}$}
The unitary $\tilde{T}$ can be treated as a state preparation process controlled by the first register to mark its row index. For simplicity and without loss of generality, we assume constant $s$ is a power of 2. From $|\tilde{j}\rangle$ and concerning the first, third and fourth register, we perform Hadamard on the third register to create the superposition

\begin{equation}
    \frac{1}{\sqrt{s}} \sum_{l=0}^{s-1}|j, l, 0\rangle_{j,k,k_c} .
\end{equation}

We make a tiny modification to the given oracle by defining $O_A'|j, l, z\rangle = |j, l, z\oplus a_{i,l}\rangle$. Note that $O_A'$ and $O_A$ are equivalent because $O_A' = O_s^\dagger O_A O_s$. The benefit is that $O_A'$ can be directly implemented with one QRAM query in the CSC format. By applying $O_A'$ and $O_s$ we obtain
\begin{equation}
    \frac{1}{\sqrt{s}} \sum_{k: A_{j k} \neq 0}|j, k, 0\rangle|A_{jk}\rangle .
\end{equation}

\begin{figure*}[t]
    \centering
    \includegraphics[width=.8\linewidth]{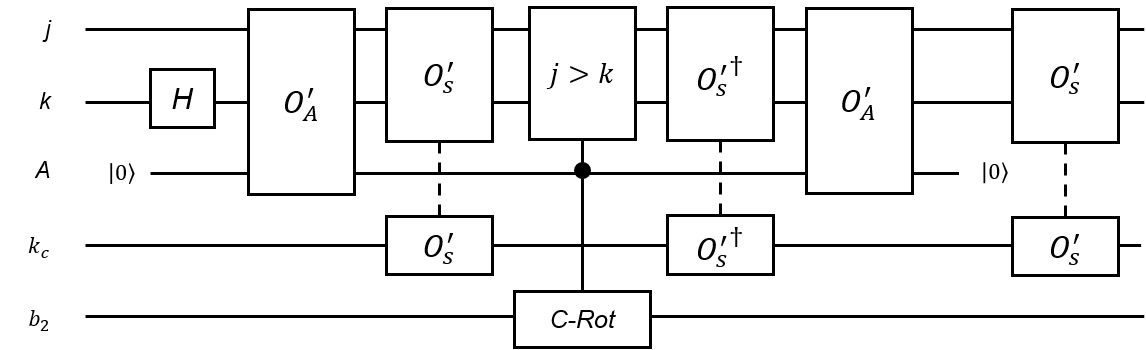}
    \caption{The quantum circuit of $\tilde{T}$. A dashed line connects one module applied on separated registers.}
    \label{fig:T}
\end{figure*}

Finally, we perform a conditional rotation\footnote{Here, the conditional rotation is controlled by the row and column indices to deal with the negative element.} and uncompute $|A_{jk}\rangle$, we can prepare the wanted $|\psi_j\rangle$. To summarize, we show the circuit in Fig.~\ref{fig:T} and the pseudocode of implementing $\tilde{T}$ in Alg.~\ref{alg:T}.

\begin{figure}[h]
    \centering
    \includegraphics[width=\linewidth]{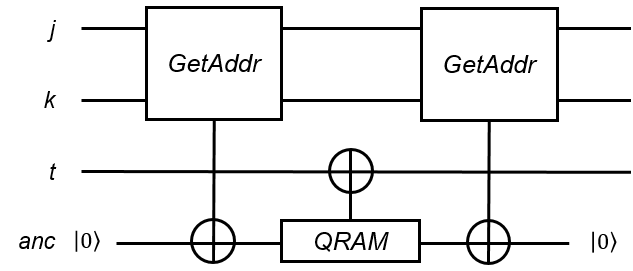}
    \caption{The quantum circuit of $O_A'$.}
    \label{fig:oa}
\end{figure}

\paragraph{Implementation of $O_A'$}
$O_A'$ makes access to the QRAM and queries the matrix element with the given input. The QRAM can be written as the following unitary
$$
U_{\rm QRAM}|i\rangle|j\rangle \mapsto |i\rangle|j\oplus d[i]\rangle.
$$
where $i$ is the address, $d[i]$ its corresponding data entry and $\oplus$ denotes the bitwise XOR. Implementing $O_A'$ is straightforward. First, we compute the address from two input registers and the offset, then we call the QRAM and finally uncompute the first step. The circuit is shown in Fig.~\ref{fig:oa}.

\paragraph{Implementation of the quantum walk}


With the modified $O_s$ based on the QBS, now the quantum walk is defined on six registers. Four registers are required by the original quantum walk theory, naming as $j$, $b_1$, $k$, and $b_2$, each has n, 1, n, and 1 qubit correspondingly. Besides these four, we also extend two more $n$ qubits registers, naming as $j_c$ and $k_c$. 

Then we define $|\tilde{j}\rangle=|0,j,0,0,0,0\rangle_{j_c, j,b_1,k_c,k,b_2}$. Also, we expand the $|\psi_j\rangle$ to $|\tilde{\psi}_j\rangle$ by filling zero state similarly. After expansion, we implement a unitary $\tilde{T}$ which prepares from $|\tilde{j}\rangle$ to $|\tilde{\psi}_j\rangle$, which corresponds to the process where the extended registers begin from $|0\rangle$ and return to $|0\rangle$ after the preparation finishes, that is
\begin{equation}
    \tilde{T} = \sum_{j \in[N]}|\tilde{\psi}_j\rangle\langle \tilde{j}|+\sum_{j^{\perp}}|\tilde\psi_j^\perp\rangle\langle\tilde{j}^\perp|.
\end{equation}
Note that 
\begin{equation}\label{eqn:impl}
 \begin{aligned}
&\tilde{T}\left( 2\sum _{j\in [N]} |\tilde{j} \rangle \langle \tilde{j} |-\mathbbm{1}\right)\tilde{T}^{\dagger } \\
=&(2TT^{\dagger } -\mathbbm{1} )_j\otimes |0\rangle\langle 0|_{j_c,b_1,k_c,k,b_2} \\ 
&+\sum _{m\neq 0,n\neq 0} \bm{V}^{mn}_{j} \otimes |m\rangle \langle n|_{j_c,b_1,k_c,k,b_2}\\
&=\begin{pmatrix}
2TT^{\dagger } -\mathbbm{1} &\\
 & *
\end{pmatrix},
\end{aligned}
\end{equation}
forms a block encoding of the wanted operator $2TT^\dagger-\mathbbm{1}$ where $2TT^{\dagger } -\mathbbm{1} $ is defined in the orthogonal subspace of $|0\rangle\langle 0|$. $2 \sum_{j \in[N]}|\tilde{j}\rangle\langle\tilde{j}|-\mathbbm{1}$ will conditionally occur a phase flip if the state is not $|\tilde{j}\rangle$, and we name it after $P$.

To prove that $\tilde{T}$ can effectively represent the $T$, note that
\begin{equation}
   \begin{aligned}
    \tilde{T}S\tilde{T}^\dagger = \begin{pmatrix}
    A/s & *\\
    * & *
    \end{pmatrix}
   \end{aligned}    
\end{equation}
forms a block encoding of $H=A/s$ where $S$ swaps $j_c,j,b_1$ with $k_c,k,b_2$. This block encoding form is compatible with the theoretical analysis of the walk operator, and we display it in the Supplementary Material. Roughly speaking, we extend the matrix size from $N$ to $N^2$, where the extended registers $j_c$ and $k_c$ are now considered as the higher digits of the row and column indices, namely $\overline{j_cj},\overline{k_ck}$, and $A$ exists in the block where $j_c=k_c=0$.

\begin{figure}[t]
    \centering
    \includegraphics[width=\linewidth]{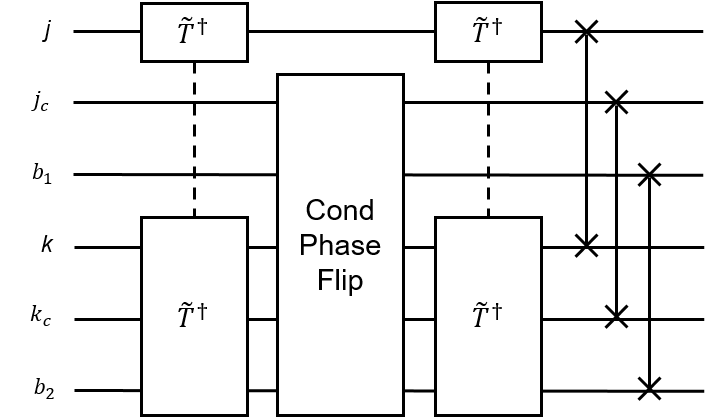}
    \caption{The quantum circuit of the walk operator $W$. A dashed line connects one module applied on separated registers.}
    \label{fig:w}
\end{figure}

Now we have a unitary implementation of the walk operator $W=S\tilde{T}P\tilde{T}^\dagger$. To summarize, we show the circuit in Fig.~\ref{fig:w}. 

\paragraph{Implementation of the CKS quantum linear solver}

The CKS quantum linear solver is based on the quantum walk, and is nearly optimal for all parameters. In theory, walking $n$ steps correspond to the n-th order Chebyshev polynomial of the first kind, namely
\begin{equation}
T^\dag W^nT|\psi\rangle|0\rangle = \mathcal{T}_n(A/d)|\psi\rangle|0\rangle + |\Psi^\perp\rangle.
\end{equation}
$|\Psi^\perp\rangle$ is some garbage output where the second flag register does not output $|0\rangle$. Additionally, $A^{-1}$ can be $O(\epsilon)$ approximated by $f(A)$ where
\begin{equation}\label{eqn:fx}
\begin{aligned}
    f(x)&=\frac{1-\left(1-x^{2}\right)^{b}}{x} \\
    &=4 \sum_{j=0}^{j_0}(-1)^{j}\left[\frac{\sum_{i=j+1}^{b}\begin{pmatrix}
2b \\
b+i
\end{pmatrix}}{2^{2 b}}\right] \mathcal{T}_{2 j+1}(x),
\end{aligned}
\end{equation}
and eigenvalues of $A$ lie in $[-1,-1 / \kappa] \cup[1 / \kappa, 1]$, $\kappa$ the condition number of matrix $A$, $j_0=\sqrt{b\log(4b/\epsilon)}$ and $b=\kappa^2\log(\kappa/\epsilon)$. Hence, $f(A)|b\rangle$ is the weighted sum of a series of $\mathcal{T}_{2 j+1}(A)|b\rangle$. Using the linear combination of unitaries (LCU) technique, we can compute the weighted sum of different orders of the Chebyshev polynomial to implement $f(A)|b\rangle$, which is $\epsilon$-close to the $A^{-1}|b\rangle$.

\section{Classical simulation of the quantum walk program\label{sec:classical}}


To validate the implementation, we run our quantum walk program on a quantum circuit simulator. The quantum circuit simulator is not built on any existing quantum programming language or simulation program because the simulation here often requires more than hundreds of qubits, exceeding the capabilities of nearly all well-known existing algorithms. We build a novel quantum circuit simulator, which only stores nonzero entries in a quantum state, and each quantum state represents a series of quantum registers. 




\subsection{Classical simulation of the quantum circuit on the quantum register level}

\paragraph{Main idea}
The classical simulation of the quantum circuit is a computationally hard task. This is generally because the representation of the Hilbert space grows exponentially with the number of qubits. This property usually forbids the simulation of over 100 qubits, which also limits the ability to conduct numerical research on large-scale quantum algorithms, such as the quantum walk used in this paper. To overcome this difficulty, we make two changes to the current classical simulation algorithms. 

First, we observe that in most of the quantum algorithms, the nonzero entries of the quantum state do not always fill the entire Hilbert space. When there are $n$ working qubits, the quantum state is often not $2^n$. In the quantum walk, the maximum number of nonzero entries usually grows polynomially with the dimension of the matrix and the sparsity number. This property allows the compression of the quantum state and greatly reduces the storage space. Also, a sparse state is more efficient to compute, where the time complexity is also polynomial dependent on the number of nonzero entries. 

Second, the basic unit of the computation is raised to the quantum register level instead of the qubit level. In large-scale quantum algorithms, quantum arithmetics are performed on the register level. When the error is small enough to support running such algorithms, a quantum arithmetic operation can be packed and simulated as a whole. This also benefits the programming of quantum algorithms, where a more natural description of the quantum program is allowed, such as instructions like ``ADD'', ``SUB'', or ``MUL''.

\begin{figure*}[t]
    \includegraphics[width=\linewidth]{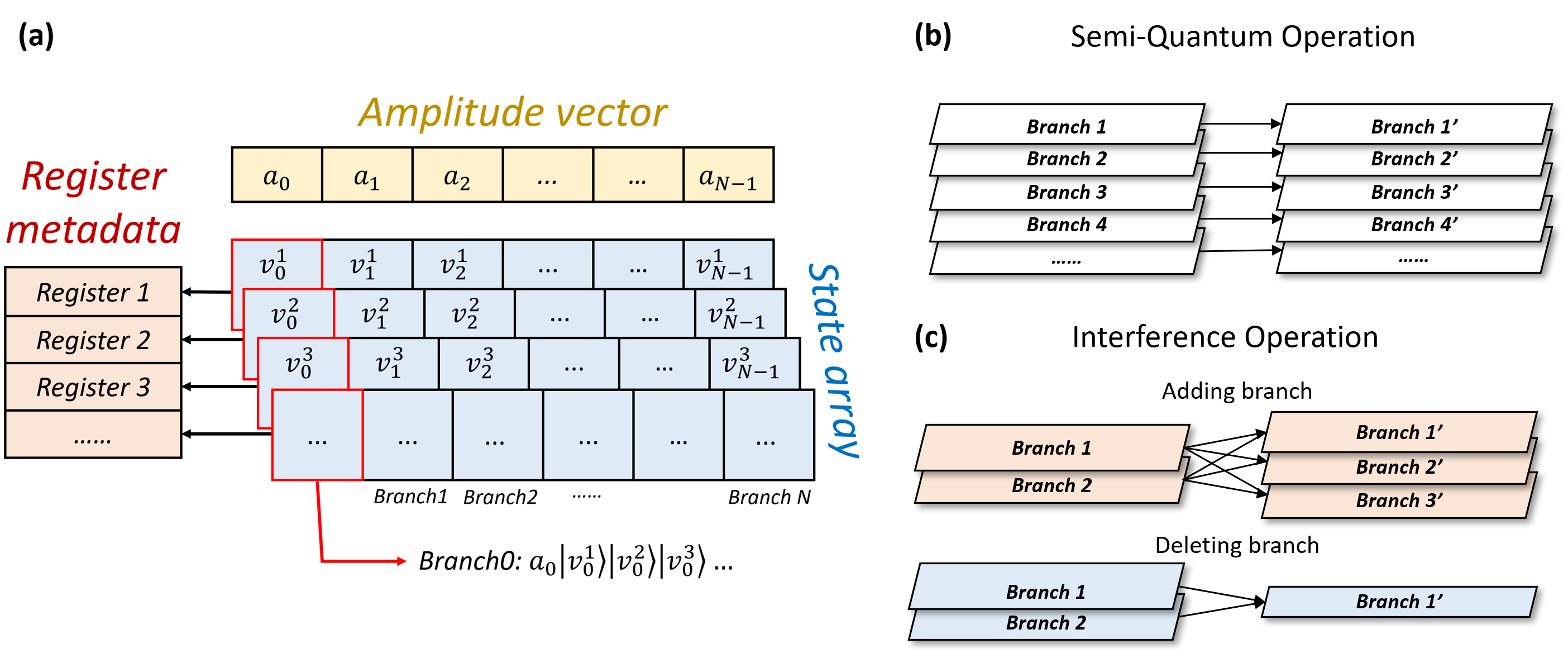}
    \caption{Schematic diagrams of the sparse state simulation. (a) The data structure of the state representation. Here, to store a quantum state with $K$ registers and $N$ amplitudes, three parts of data are stored. The \textit{amplitude vector} stores $N$ complex number to represent the amplitude of each state, namely the amplitude of a \textit{branch}. The \textit{register metadata} stores the metadata for $K$ registers. And the \textit{state array} is a $K\times N$ array that stores the value of registers of all branches. (b) The semi-quantum operation is to compute each branch independently, including quantum arithmetics, phase flipping, and QRAM operations. (c) Interference operations will add branches or delete branches. This includes the Hadamard operation, Fourier transforms, etc.}
    \label{fig:sparse_simulation_app}
\end{figure*}

The idea of leveraging the sparsity of the quantum state is not initially proposed in our work, but applying the sparse simulation on the register level can dramatically increase the scale of the quantum algorithm that a classical computer can simulate. The programming and simulation of the quantum walk are based on these two changes, so we design a novel simulation program that supports these ideas. Suppose a quantum state that can be written as 
\begin{equation}
    |\psi\rangle = \sum_i a_i|v^1_i\rangle|v^2_i\rangle|v^3_i\rangle...,
\end{equation}
where $v^j_i$ is a computational basis state of register $j$ in the branch $i$. Define the number of registers as $K$, and the number of branches as $N$. This quantum state can be represented by a $K\times N$ array that stores every $v_i^j$, and a $N$-sized vector that stores every $a_i$. In Fig~\ref{fig:sparse_simulation_app}(a), a schematic of the data structure is shown. We name the $K\times N$ array as the \textit{state array}, and the vector that stores the amplitude as the \textit{amplitude vector}.

A state array's element can be stored as a plain binary string whose length corresponds to the number of qubits of the register. In practice, all elements are represented by a fixed 64-bit binary string, and the \textit{register metadata} is recorded to track the size of all registers. Creating a new register is to append a row to the state array as well as the corresponding metadata, and removing a register is to remove the row. 


When performing arithmetic computations, the data in registers are often processed with some data type. Generally, this data structure does not limit any certain kind of data type, but we still provide some typical types (integer, unsigned integer, fixed-point number, boolean, etc.) and enable the runtime type check.

The operations can be divided into two types, as demonstrated in Fig.~\ref{fig:sparse_simulation_app}(b) and (c). The first type is the semi-quantum operation, which processes each branch in parallel and will not change the number of branches, such as quantum arithmetics and QRAM. The second type is the interference operation. This operation will cause quantum interference among several branches, where some branches may be added or deleted. In the quantum walk, Hadamard and conditional rotation is of this type. Detailed descriptions of the simulation algorithm are listed in the Supplementary Material.

\paragraph{Classical simulation of the quantum arithmetic}

A quantum arithmetic operation is usually a semi-quantum process that satisfies the following properties:
\begin{itemize}
    \item All branches are computed independently (in parallel). No branch will be created or deleted.
    \item It is applied on several registers, and the other registers remain unchanged.
    \item It is usually reversible and creates no garbage. If not, then we can extend the input register set with the garbage registers to make it reversible.
\end{itemize}

The register-level simulation of the quantum arithmetic basically follows these steps. First, iterate over all branches, and for each branch, extract the contents (binary strings) of the input registers. Then use type conversion to process the binary strings into certain data types and directly perform the computation. Finally, change the content in the registers.

The most vital problem is that one should guarantee the change of contents of the registers is reversible. Here we introduce a simple protocol for out-of-place arithmetics. Suppose a classical function $y=f(x_1,x_2,...,x_n)$ where $x_i$ is the input and $y$ is the output, and the following operation $U$ is guaranteed reversible:
\begin{equation}
    U_f|x_1\rangle|x_2\rangle...|x_n\rangle|z\rangle = |x_1\rangle|x_2\rangle...|x_n\rangle|z\oplus y\rangle,
\end{equation}
where $\oplus$ is the bitwise XOR. With this protocol, simulating the quantum arithmetic only has to extract all input $x_i$, compute $y$, and finally change the content in the output register. 

For the in-place operation, there are two possible methods. One is to manually compute the mapping and guarantee it to be a bijection mapping. Another is to implement the out-of-place version of the function and the inverse function. For example, if the output of $y$ overwrites $x_1$, such that 
\begin{equation}
    U_{f}^{\mathrm{in-place}}|x_1\rangle|x_2\rangle...|x_n\rangle = |y\rangle|x_2\rangle...|x_n\rangle.
\end{equation}
Then it is equivalent to implementing $U_f$ and $U_{f^{-1}}$ such that 
\begin{equation}
    U_{f^{-1}}|y\rangle|x_2\rangle...|x_n\rangle|z\rangle = |y\rangle|x_2\rangle...|x_n\rangle|z\oplus x_1\rangle,
\end{equation}
where $x_1=f^{-1}(y,x_2,...,x_n)$.

\paragraph{Classical simulation of the quantum random access memory}

The simulation of the quantum random access memory (QRAM) is similar to the simulation of quantum arithmetic operations. QRAM can implement a unitary $U_{\mathrm{QRAM}}$ such that
\begin{equation}
    U_{\mathrm{QRAM}}|i\rangle_A|z\rangle_D = |i\rangle_A|z\oplus d_i\rangle_D.
\end{equation}
Here, QRAM manages a vector of memory entries $\vec{d}$ and has two variables to control the input: the address length and the word length. The address length $n$ is the size of the address register $|\cdot\rangle_A$, and the word length is the size of data register $|\cdot\rangle_D$. The word length is also the effective size of a memory entry $d_i$. When the noise is not considered, the simulation of QRAM is to access the classical memory with the content of the address register.

\paragraph{Classical simulation of the interference operation}
The simulation of the interference operation includes three major steps:
\begin{enumerate}
    \item Grouping;
    \item Computing;
    \item Clearing zero elements.
\end{enumerate}

Suppose the interference operation is acted on a set of input registers, and we call remained registers \textit{idle registers}. We should notice that any two branches are coherent if and only if the values of all idle registers are the same. The first step is to sort all branches, then split branches into groups in which all registers have the same value except input registers. 

In each group, we can just calculate a small number of branches. If a group has branches the same number as the dimension of the operation (for example, a Hadamard on $k$-bit register, and a group with $2^k$ branches), then this group can be computed in-place. Otherwise, we may append some branches and perform operations on the full size.

After the computation step, some branches may have zero amplitude due to the effect of quantum interference, and these branches will be removed from the branch list to reduce memory consumption.

\subsection{Numerical results}

We randomly generate positive-valued symmetric band matrices, which is a typical type of matrix arising from realistic world tasks. Two variables control the generation of the matrix: the number of rows $N$, and the number of bands. Each matrix element is a fixed point number within $[0,1)$, and the number of digit is called \textit{word length}.

Before starting the solver program, we have to preprocess those sparse matrices with the following steps:

\begin{enumerate}
    \item If $\lVert A\rVert_{\rm max} > 1$, rescale it to be 1;
    \item Convert the sparse matrix to CSC format;
    \item Count the maximum number of nonzeros in a row, recorded as $s$;
    \item Fill zeros if any row does not have $s$ nonzeros;
    \item Calculate the minimum eigenvalue of $A/s$ as $\lambda_{\rm min}$, then $\kappa = 1/\lambda_{\rm min}$.
\end{enumerate}

\begin{table*}[t]
\caption{Experimental settings}
\vspace{0.1cm}
\centering
\begin{threeparttable}
\begin{tabular}{ccc|ccc|ccc}
\toprule[2pt]
\multicolumn{3}{c|}{Metadata} & \multicolumn{3}{c|}{Related variables} & \multicolumn{3}{c}{Simulation resources}          \\ 
\midrule[1pt]
  \begin{tabular}[c]{@{}c@{}}Matrix \\ name\end{tabular} &
  \begin{tabular}[c]{@{}c@{}}Row \\ size\end{tabular} &
  \begin{tabular}[c]{@{}c@{}}Word \\ length\end{tabular} &
  \begin{tabular}[c]{@{}c@{}}Condition\\ number $\kappa$\tnote{a}\end{tabular} &
  \begin{tabular}[c]{@{}c@{}}Sparsity\\ $s$\tnote{b}\end{tabular} &
  $j_0$\tnote{c} &
  Qubit number\tnote{d} &
  \begin{tabular}[c]{@{}c@{}}Average time\\ per walk step\end{tabular} &
  \begin{tabular}[c]{@{}c@{}}Memory usage\end{tabular}\\
\midrule[1pt]
 A &  16  & 8  & $1.0\times 10^2$ & 8  & 1532 & 165 & 1.1 ms & 162 KB \\
 B &  128 & 8  & $6.6\times 10^2$ & 16 & 11824 & 245 & 12 ms & 1.54 MB \\
 C & 1024 & 8  & $5.0\times 10^3$ & 32 & 106122 & 365 & 1.2 s & 49.4 MB \\ 
 D & 8192 & 16  & - & 32 & - & 428 & 14 s & 859 MB \\ 
 E & 16384 & 32  & - & 64 & - & 582 & 1.1 min & 1.66 GB \\ 
 \bottomrule[2pt]
\end{tabular}
\begin{tablenotes}    
\footnotesize               
\item[a] The inverse of the minimal absolute eigenvalue of $A/s$.
\item[b] The number of nonzero entries in a row is rounded up to $2^{\lceil\log \tilde{s}\rceil}$, where $\tilde{s}=2\times \mathrm{Bandwidth}+1$.
\item[c] Related to the order of summation in Eqn.~(\ref{eqn:fx}).
\item[d] Automatically counted by the simulator, which is the maximum number of working qubits.
\end{tablenotes}
\end{threeparttable}
\label{tab:experiment}
\end{table*}

Then we compute the output state $|\tau_j\rangle$ for all $j\leq j_0$ where
\begin{equation}
    |\tau_j\rangle = \mathcal{T}_{2j+1}(A/s)|b\rangle|0\rangle_{\mathrm{flag}}+|\tau^\perp\rangle.
\end{equation}
To avoid repetition, we compute them iteratively, that is $|\tau_j\rangle = \tilde{T}^\dagger W^2\tilde{T}|\tau_{j-1}\rangle$. The output state at step $j$ is defined as
\begin{equation}
    |T_j\rangle = \sum_{k=0}^{j} a_{k} |\tau_{k}\rangle.
\end{equation}

The successful rate at step $j$, denoted as $p_j$, is defined by the probability that the flag register of $|T_j\rangle$ is measured to $|0\rangle$, namely 
\begin{equation}
    p_j = \operatorname{Tr}\left( \left(I\otimes |0\rangle\langle 0|\right)|T_j\rangle\langle T_j|\right).
\end{equation}
After measurement, the state in the work register is used to evaluate the fidelity of the target state $|x\rangle = A^{-1}|b\rangle/\|A^{-1}|b\rangle\|$, that is 
\begin{equation}
    F_j = \left| \langle x,0|T_j\rangle\right|^2.
\end{equation}
The final result of the CKS solver is the state at $j=j_0$, which has the successful rate $p_{\mathrm{succ}}=p_{j_0}$ and fidelity $F = F_{j_0}$. 

\begin{figure*}[t]
    \centering
    \includegraphics[width=\linewidth]{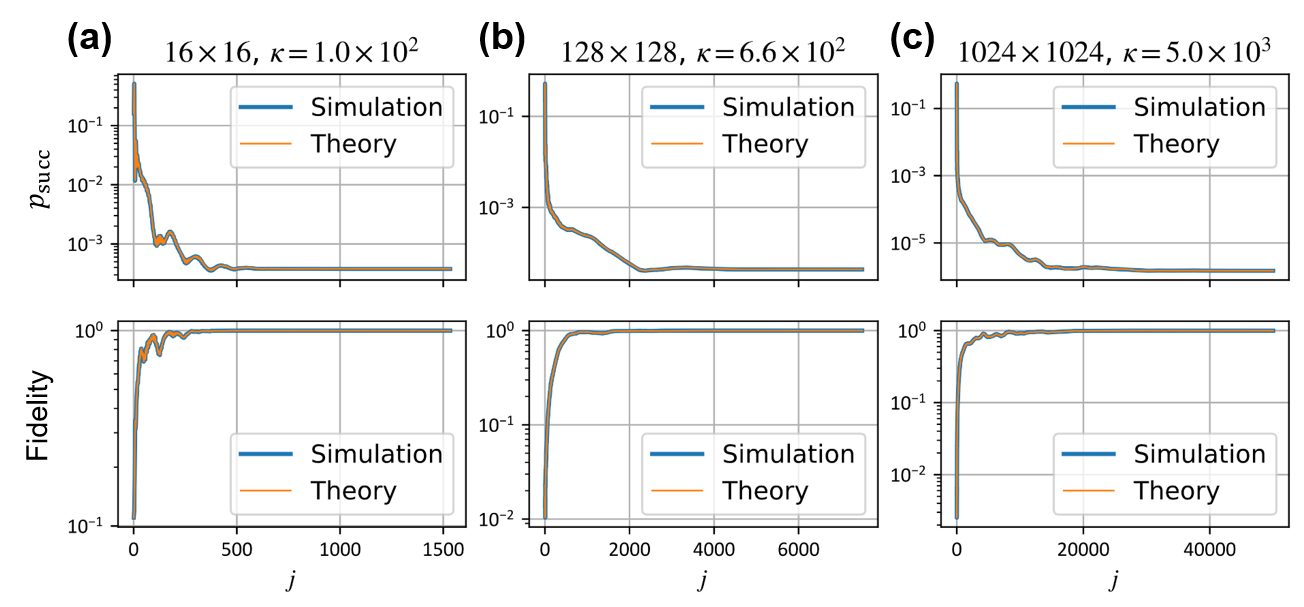}
    \caption{The results of classical simulations conducted on matrices of varying sizes and condition numbers. Each subfigure represents a specific matrix size and condition number combination, with the matrix size and condition number identified in the subtitle. The top subfigure displays the success rate $p_{\rm succ}$ in generating the desired output state $|T_j\rangle$ at each step $j$, while the bottom subfigure shows the fidelity to the target output. Note that the maximum value of $j$ depicted in the plots does not always correspond to the final step $j_0$, but rather to a point at which the computation nearly converges. Both the simulation output and the theoretical result are demonstrated and compared in the same figure. The subfigures correspond to matrices of sizes and condition numbers (a) 16 and $\kappa=1.0\times 10^2$, (b) 128 and $\kappa=6.6\times 10^2$, and (c) 1024 and $\kappa=5.0\times 10^3$.}
    \label{fig:numerical result}
\end{figure*}  

All simulations are conducted on a single core of Intel Xeon E5-2680. We generate five matrices, whose row sizes ranged from 16 to 16384, whose metadata, related variables, and simulation resources are listed in Table~\ref{tab:experiment}. The test method is the linear solver on the matrix A, B, and C with given input $\vec{b}$ as a random uniform vector, and their results are shown in Fig.~\ref{fig:numerical result}. The fidelity and the successful rate versus the iteration step $j$ are plotted. In all three matrices, the curves match perfectly with the theory, which proves the correctness of our program implementation. The plotted $j$ is less than expected $j_0$ because the computing converges and results do not change since then. For matrices D and E, the test method is to perform 50 steps of the quantum walk and their results also match perfectly with the theory.

To summarize, we calculate the quantum walk on $16384\times 16384$ matrix with $s=64$ with 1.1 minutes per step and calculate the linear solving on $1024\times 1024$ matrix with $s=32$ with 1.2 seconds per step. The number of walking steps for the $1024\times 1024$ is 212245 ($=2j_0+1$), and the total time is approximately 70 hours.

The qubit number shown in the table is counted automatically by the simulator. We exclude qubits that are not used in the computation even though they are allocated. In the quantum walk, the maximum number of qubits is counted when it reaches the deepest recursion of the quantum binary search. We provide a way to estimate the number of qubits in the Supplementary Material. But we also have to mention that here the qubit number is not a good measure of the problem size. Instead, the maximum number of nonzero state components is more important. For reference, the average time per walk step and the maximum memory usage is also provided in the table. The time and memory usage affects greatly by the row size and the sparsity of the matrix.

\begin{figure}[htbp]
\begin{overpic}[width=0.9\linewidth]{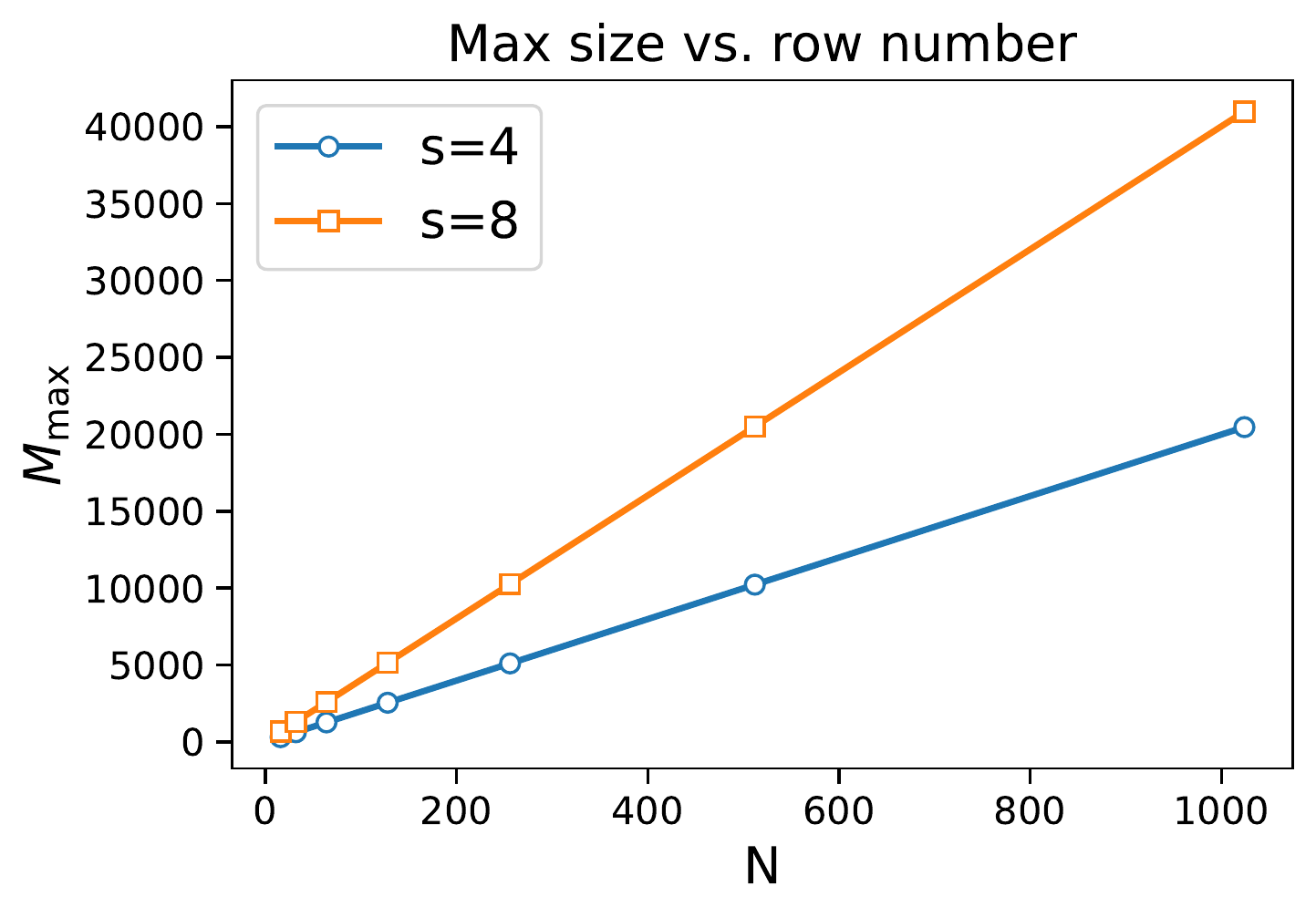}
  \put(1,65){\textbf{(a)}}
\end{overpic}
\begin{overpic}[width=0.85\linewidth]{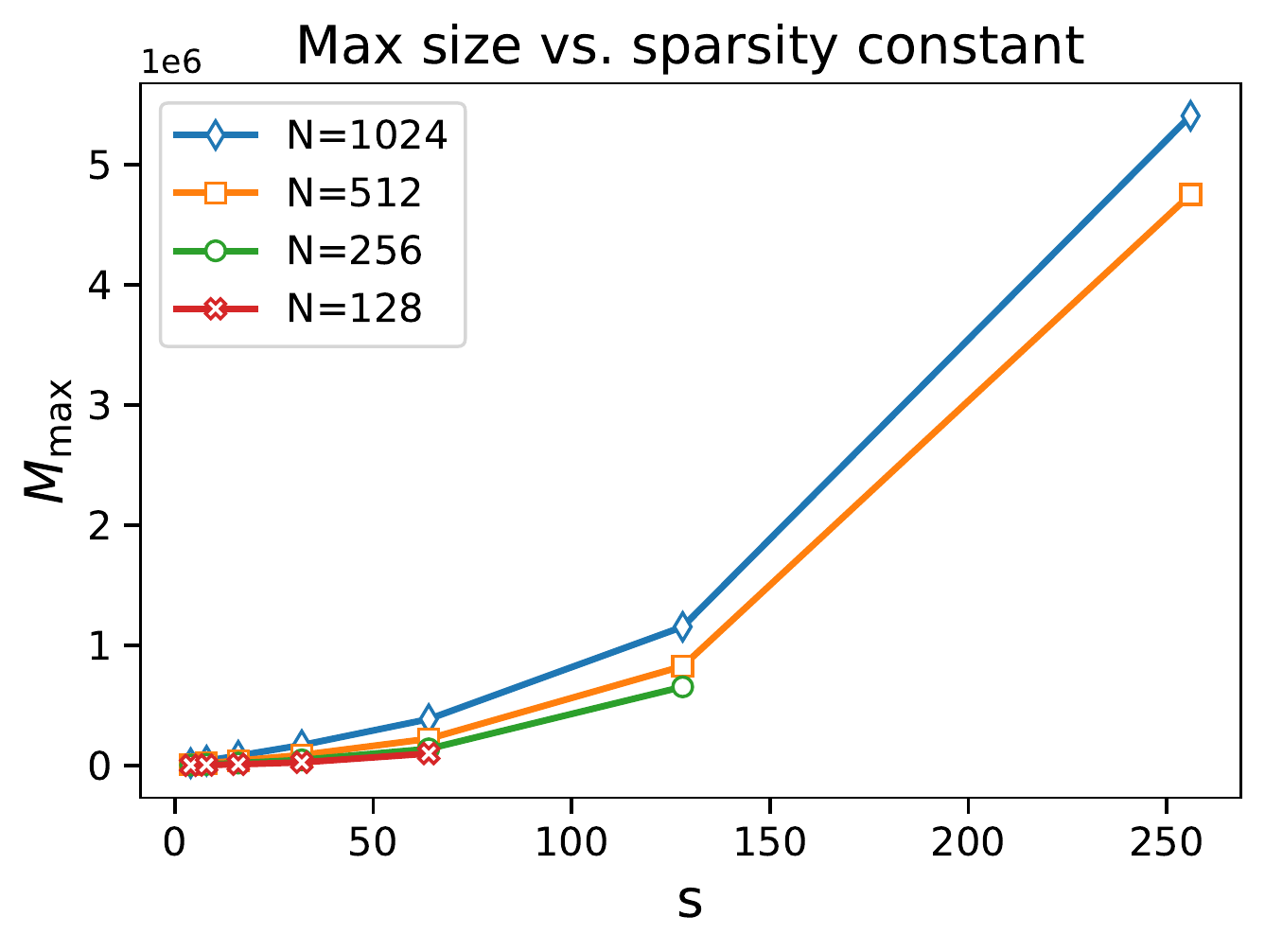}
  \put(1,68){\textbf{(b)}}
  \end{overpic}
\begin{overpic}[width=0.85\linewidth]{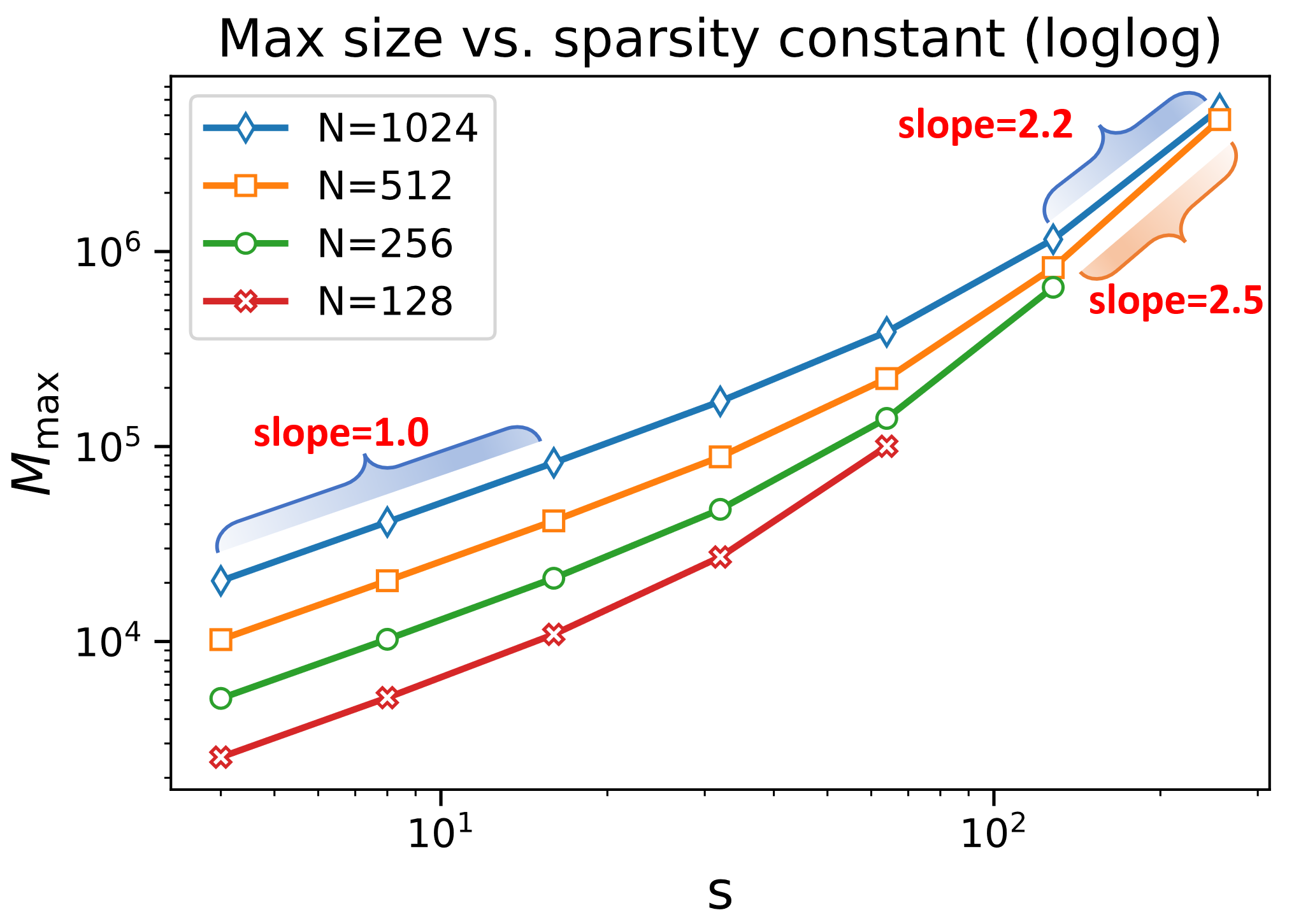}
  \put(1,65){\textbf{(c)}}
\end{overpic}
\caption{Test results of the maximum system size $M_{\mathrm{max}}$ versus the row number $N$ or sparsity constant $s$. (a) Max size versus row number. (b) Max size versus sparsity constant. (c) Max size versus sparsity constant (log-log).}

\label{fig:max_size_result}
\end{figure}
\subsection{Complexity of the classical simulation}

\subsubsection{Qubit and register usage}

In the main text, we list qubit numbers that are used in the quantum walk on different sizes of matrices, which are automatically counted by the quantum circuit simulator. During the computation, the maximum number of working registers and working qubits are maintained. When a register is allocated, the simulator will update the number of working registers if the system uses more registers than before. When a register is deallocated, the number will not shrink, which implies the recycling of the free space. The number of working qubits is the sum of the qubit number of each working register.

Here we also show how to theoretically estimate the qubit number in the quantum walk. It is in the quantum binary search when the system requires the biggest number of registers. Each QBS loop cycle will push four interval variables: \textit{mid}, \textit{midVal}, \textit{compareLess}, and \textit{compareEqual}, which have $n$, $k$, 1, and 1 qubit number correspondingly. Here $n$ and $k$ are the sizes of the address and data which are the QRAM's input, where $n=\log N + 1$ and $k$ is a set value. The number of the loop cycles is $\log s+1$ in order to locate every possible input. Then the qubit number used as the ancilla in the QBS is $(\log s+1)(n+k+2)$. The remaining part is the six registers used in the quantum walk, which have $4n+2$ qubits. So that in total, the number of qubits is
\begin{equation}
    (n+k+2)\log s+5n+k+4.
\end{equation}

\subsubsection{The space complexity of simulating the quantum walk}

The space complexity of this method can be written directly as $O(K_{\mathrm{max}}M_{\mathrm{max}})$ where $K_{\mathrm{max}}$ is the maximum number of working registers, and $M_{\mathrm{max}}$ is the maximum number of branches. In the quantum walk scenario, $K_{\mathrm{max}}$ is logarithmic dependent on the problem size. In the quantum binary search, a constant number of ancilla registers will be added for each loop cycle, and the number of the loop cycle is $O(\log s)$. Except for this, the number of working registers is always a constant number. Therefore, $K_{\mathrm{max}}=O(\log s)$.

Next we estimate $M_{\mathrm{max}}$. The nonzero entries are created during the preparation of $|\tilde{\psi}_j\rangle$. In the first application of $\tilde{T}$ on the initial state $|b\rangle = \sum_j \alpha_j|j\rangle$, we have
\begin{equation}
  \tilde{T}|b\rangle = \sum_{j,k} \alpha_j\sqrt{A_{jk}^*} |j,k\rangle + \sqrt{1-|A_{jk}|}|j,k+N\rangle,
\end{equation}
where $M=O(Ns)$ because it includes all nonzero elements in the matrix, which is a lower bound for the $M_{\mathrm{max}}$. After the first walk step, we will produce
$$\tilde{T}^\dagger S \tilde{T}|b\rangle.$$
In each application of $\tilde{T}$, two state preparation processes are applied, where each will maximumly create $s$-fold copies of the quantum state. Therefore, we will obtain at most $O(Ns^3)$ nonzeros entries.

In the theory, the quantum walk operator is a rotation on the subspace span $\left\{T|\lambda\rangle, |\perp_\lambda\rangle \right\}$, where $|\lambda\rangle$ is the eigenstate of $H$ with eigenvalue $\lambda$, and $|\perp_\lambda\rangle$ is orthogonal to $|\lambda\rangle$ in the subspace. For any input $|b\rangle = \sum_\lambda b_\lambda|\lambda\rangle$, the first walk step has already created a superposition over two states in the subspace, that is $\tilde{T}^\dagger S \tilde{T}|b\rangle = \sum_\lambda  b_\lambda
(\lambda|\lambda\rangle + \sqrt{1-\lambda^2}|\perp_\lambda\rangle)$. In the following steps, it only changes the coefficient without evidently changing the number of the nonzero entries on the computational basis. In conclusion, the maximum number of nonzero entries during the whole quantum walk process is upper-bounded by $O(Ns^3)$.



Here we plot the max size obtained from numerical experiments on various matrix configurations in Fig.~\ref{fig:max_size_result}. In subfigure (a), the max size is approximately linear-dependent on the row number $N$ in the case of $s=4$ or $s=8$. In subfigure (b), the dependence on sparsity constant $s$ is nonlinear. To further investigate the relationship, we plot them with a log-log scale in subfigure (c) and compute the slope. From this result, we estimate that $M_{\mathrm{max}}$ is $O(Ns)$ when $s$ is significantly less than $N$, and between $O(Ns^2)$ and $O(Ns^3)$ when $s$ approaches $N$.

\subsubsection{The time complexity for simulating the quantum walk}

After determining the maximum number of nonzero entries in the quantum state, now we can estimate the time complexity of the simulation.

In each quantum walking step, there are two interference operations: one is the state preparation on the register $k$ to create superposition, and another is to uncompute this preparation to achieve a reflection. Except for these two interference operations, the remaining are all semi-quantum operations.

For the semi-quantum operations, the time complexity can be computed by the number of nonzero entries $M$ and the number of the operations, which is $O(\log s)$, mainly contributed by the quantum binary search. Considering the maximum number of nonzero entries estimated in the previous text, the time complexity for the semi-quantum operations is $O(Ns^3\log s)$.

For the interference operation, the time complexity is mainly contributed by the sorting of the quantum state that creates groups. Sorting requires $O(M\log M)$ time complexity, and the other part is proportional to $M$. Therefore, the time complexity should be $O(Ns^3\log {Ns})$.

Adding up these two parts, we obtain an estimated time complexity for simulating each walking step, that is $O(Ns\operatorname{polylog}(Ns))$. Finally, we consider the number of the walk step required for this simulation $j_0$, which is approximately proportional to $\kappa$. In total, the time complexity for conducting the simulation for the entire quantum linear solver is 
\begin{equation}
    O(N\kappa\operatorname{poly}(s)\log N ).
\end{equation}

A common knowledge is that simulating quantum computing will introduce many extra overheads, and the simulation ability is much weaker than using a classical algorithm and solving it directly. Surprisingly, the simulation of the quantum walk approaches the complexity of the classical computation, such as the conjugate-gradient method with $O(N\kappa s)$ time complexity. For Other quantum circuit simulation methods, the number of qubits and depth of the circuit can often be the restriction and they can hardly achieve a linear dependence on the size of the matrix. In our view, the improvement is mainly caused by only tracking the nonzero entries in the quantum state, and the size of the quantum state is often proportional to the intrinsic complexity of the problem itself.



\section{Summary and Outlook\label{sec:summary}}

We believe that programming the quantum algorithm, as well as simulating it with realistic data, is one of the most convincing and concrete ways to study quantum algorithms. However, there is always a gap between the quantum algorithm and an executable quantum program. How to decompose each step of the quantum algorithm into a set of basic operations is the key to filling the gap.

In this article, we developed a series of techniques to program the quantum walk on a sparse matrix stored in the QRAM, including a unitary quantum binary search without side-effect and the implementation of the quantum walk on the expanded space. We execute this program on a quantum circuit simulator, test it on several matrices, and compare it with the theory.


All the techniques developed in this paper can also be used in the programming of other quantum algorithms, e.g. implementing a quantum while loop and dynamical allocation and deallocation of ancilla registers. Because the quantum walk operator also naturally generates the block encoding of a sparse matrix, we can expect the implementation and simulation of the quantum linear algebra\cite{LinearAlg} and grand unification of quantum algorithms\cite{grand}.

We would like to conduct two future works related to this paper. The first is to add errors to the simulation. The error should be the most critical factor that affects quantum computation. In fault-tolerant quantum computing, the required error rate determines the amount of error correction resources. A noisy simulation could be a key to discovering the minimum resource that a quantum computer requires to achieve quantum speedup in some real-world problems. The second is to estimate the execution time and T-count. It is possible to estimate these via a validated and detailed program implementation, and an automatic tool for the resource and time estimation of a large-scale quantum algorithm might be created in the future.


\section*{Acknowledgement}
This work was supported by the National Natural Science Foundation of China (Grant No. 12034018), and Innovation Program for Quantum Science and Technology No. 2021ZD0302300.


\appendix


\begin{table*}[htbp]
\centering
\caption{Checklist for all basic operations used in this paper}
\vspace{0.1cm}
\begin{threeparttable}
\begin{tabular}{cccc}
\toprule[2pt]
Operation name & Operand type &  Reference & Related process \\ 
\midrule[1pt]
Addition          &  Int & \cite{QArith1} & $O_A'$, $O_s$\\
In-place Addition &  Int & \cite{QArith1} & $O_A'$, $O_s$ \\
Multiplication    &  Int & \cite{QArith1} & $O_A'$, $O_s$  \\
Comparator        &  Int & \cite{QArith1} & $\mathrm{QBS}$\\
Square root       &  Fixed point    &  \cite{QArith4}  & Conditional rotation \\
Arccos            &  Fixed point    &  \cite{QArith2}  & Conditional rotation \\
Conditional rotation & Fixed point &  \cite{QArith3}  & $\tilde{T}$ \\
Hadamard  & Int & \cite{QCQIa} & $\tilde{T}$\\
Controlled Phase Flip  &  Int & \cite{QCQIa} & Quantum walk  \\
Swap  &  Any & \cite{QCQIa} & Quantum walk  \\
\bottomrule[2pt]
\end{tabular}
\end{threeparttable}
\label{tab:arithmetic}
\end{table*}
\section{Quantum arithmetics used in this paper}
Quantum arithmetics are often regarded as the circuit implementations of arithmetic logic units (ALU) in the quantum computer, including addition, subtraction, multiplication, square root, etc. Quantum arithmetics are semi-quantum circuits, which reversibly implement classical logic computations, and are very essential in many quantum algorithms like Shor's factorization. Many previous works have introduced the implementation and optimization of some quantum arithmetics.

Different from many other works which explicitly implement all operations in the quantum circuit form, this paper terminates at some ``basic operations'' including those quantum arithmetics. To ensure these operations can be fully decomposed into quantum gate level, we list all of them in Table~\ref{tab:arithmetic}.



\begin{figure*}[htbp]
    \centering
    \includegraphics[width=0.75\textwidth]{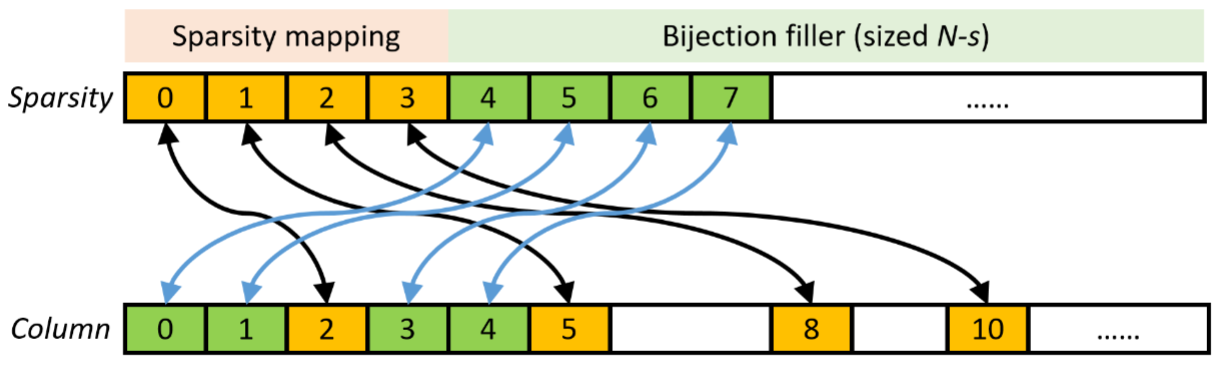}
    \caption{An intuitive bijection mapping between any $0\leq k<N$ and $0\leq l<N$. }
    \label{fig:bijection_simple}
\end{figure*}
\section{The intuitive implementation of the $O_s$\label{app:intuitive}}
In this section, we introduce how to implement the original $O_s$ completely in place without causing any side effects. First, we restate the definitions: the dimension of the matrix is $N$; sparsity is $s$; in row $j$, the index of the $l$-th nonzero column is $k$. When the matrix is stored in CSC format, it is efficient to compute $k(j,l)$ when $0\leq l<s$. Also, the quantum binary search can compute $l(j,k)$ in $O(\log s)$ time when $k$ is a correct column index. Now the task is to create a mapping from any $0\leq k<N$ to $0\leq l<N$.

The in-place mapping from $k$ to $l$ is equivalent to defining two out-of-place operations, that is
\begin{equation}
\begin{aligned}
    &|j,k\rangle \mapsto |j,k,l(j,k)\rangle\\
    &|j,l\rangle \mapsto |j,l,k(j,l)\rangle.
\end{aligned}
\end{equation}
They can be achieved by first copying the register and then performing the in-place operation on the copied register.

In Fig.~\ref{fig:bijection_simple}, we again display an intuitive example version of $O_s$. For the given input $l$, there exist two domains: one is that $l<s$, called ``sparsity mapping'', and the other is $l\geq s$, called ``bijection filler''. The first $l$ in the bijection filler domain maps to the index of the first zero columns and so forth. This mapping can be represented by the following function

\begin{equation}
    k(j,l) = \left\{
    \begin{array}{ll}
    \tilde{k}(j,l) & 0\leq l<s\\
    l-\delta_l & s\leq l<N
    \end{array}
    \right..
\end{equation}
Here $\delta_l$ is the difference between $k$ and $l$. Even though it is straightforward to compute this difference from $k$, it is necessary to compute this difference with only the given input $l$ for reversibility. Then the algorithm is as follows.

\begin{algorithm}[H]
\caption{Find $k$ from input $l\geq s$}
\label{alg:findk}
\begin{algorithmic}[1]
\Require $l$, sparsity number $s$, the nonzero column indices $a$, and matrix dimension $N$.
\Ensure Return $k$.
\State Compute $l\gets l-s$; 
\For {i = 0 :  $s-1$}
    \If {$a[i] \geq l$}
    \State \textbf{return} $l$; 
    \Else
    \State $l\gets l+1$;
    \EndIf
\EndFor
\State  \textbf{return} $l$;
\end{algorithmic}
\end{algorithm}

This algorithm sequentially finds the corresponding $k$ belongs to which section. Therefore, the time complexity for this algorithm is $O(s)$. While this algorithm also uses a for-loop that has to be run in quantum parallel, using the method that described in the main text, we can convert it to the quantum version and obtain the space complexity $O(s)$.

\section{Pseudocode of the implementation}

Here we list the pseudocode for implementing each module of the quantum walk, including $O_A'$, $O_s'$, $\tilde{T}$, and the walk operator $W$.

\begin{algorithm}[H]
\caption{Implementation of $O_A'$}
\label{alg:OA}
\begin{algorithmic}[1]
\Require Register $j$, $k$, $t$;
\Require Sparsity constant $s$, sparsity segment's offset \textit{offset};
\Ensure None
\Comment{ QuantumArithmetic, QRAM}
\State Allocate a temporal register $\rm{anc}$;
\State Compute the address on register $\rm{anc}$ with $j$, $k$, $s$, and \textit{offset}; 
\State Apply QRAM query with address register $\rm{anc}$ and data register $t$;
\State Uncompute step 2;
\State Discard ancilla $\rm{anc}$ safely.
\end{algorithmic}
\end{algorithm}

\begin{algorithm}[H]
\caption{Implementation of $O_s'$}
\label{alg:Os}
\begin{algorithmic}[1]
\Require Register $j$, $l$, $k_c$
\Require Sparsity constant $s$, sparsity segment's offset \textit{offset};
\Ensure Quantum Binary Search (\textbf{QBS})
\Comment QuantumArithmetic (in-place Addition), QRAM, Swap
\State Compute the address on register $\rm{anc}$ with $j$, $s$, and \textit{offset}, adding in-place on $l$; 
\State Apply QRAM query with address register $l$ and data register $k_c$;
\State Apply \textbf{QBS} on $k_c$ with output register $l$;
\State Swap $l$ and $k_c$ register.
\end{algorithmic}
\end{algorithm}

\begin{algorithm}[H]
\caption{Implementation of $\tilde{T}$}
\label{alg:T}
\begin{algorithmic}[1]
\Require Register $j$, $k$, $b_2$, $k_c$
\Ensure $O_A'$ and $O_s'$
\Comment{Hadamard, ConditionalRotation}
\State Allocate temporal register $A$;
\State Apply Hadamard on register $k$;
\State Apply $O_A'$ on register $j$, $k$ and $A$;
\State Apply $O_s$ on register $j$, $k$ and $k_c$;
\State Apply conditional rotation on $b_2$ controlled by element $A$, row index $j$ and column index $k$;
\State Apply $O_s^\dagger$ on register $j$ and $k$;
\State Apply $O_A'$ on register $j$, $k$ and $t$ (uncomputation of step 2);
\State Apply $O_s$ on register $j$, $k$ and $k_c$;
\end{algorithmic}
\end{algorithm}

\begin{algorithm}[H]
\caption{Quantum walk operator}
\label{alg:qw}
\begin{algorithmic}[1]
\Require Register $j_c$, $j$, $b_1$, $k_c$, $k$, $b_2$;
\Ensure $\tilde{T}$.
\Comment{ControlPhase, Swap}
\State Apply $\tilde{T}^\dagger$ on register $j$, $k$, $b_2$, and $k_c$;
\State Apply phase flipping controlled by zero state of register $b_1$, $k$, $b_2$, $j_c$, and $k_c$;
\State Apply $\tilde{T}$ on register $j$, $k$, $b_2$, and $t$;
\State Apply swapping between register $j$ and $k$, $j_c$ and $k_c$, as well as $b_1$ and $b_2$.
\end{algorithmic}
\end{algorithm}

\section{Classical binary search}

The binary search is a classic, efficient algorithm that looks up a target in a sorted list. There are many methods to implement a binary search, such as using a while-loop or recursion. Here we use a for-loop version that has a fixed step number, which can more easily be converted to the quantum version.

Note that this binary search version only correctly outputs the position when the search target is actually in the list, which has already met the demand for implementing the quantum walk.

\renewcommand\algorithmicensure{\textbf{Ensure:}}
\begin{algorithm}[H]
\caption{Classical binary search}
\label{alg:bs}
\begin{algorithmic}[1]
\Require Sized-$s$ list $a$, the search target \textit{target}.
\Ensure The position of the target and a flag indicating whether \textit{target} is found.   
\State Allocate temporal variables \textit{left}, \textit{right}, \textit{mid}, and \textit{midVal}; 
\State \textit{left} $\gets 0$, \textit{right} $\gets$ $s$;
\For {i = 0 : $\lceil \log s\rceil$}
    \State $\textit{mid} \gets$ $ \lceil(\textit{left}+\textit{right})/2\rceil$;
    \State \textit{midVal} $\gets a[\textit{mid}]$;
    \If {midVal = target}
    \State \textbf{return} $\{\textit{mid}, \text{true}\}$;   \Comment{Target is found.}
    \EndIf
    \If {$\textit{midVal} < \textit{target}$}
    \State $\textit{left}\gets\textit{mid}$;
    \Else
    \State $\textit{right}\gets\textit{mid}$;
    \EndIf
\EndFor
\State  \textbf{return} $\{0, \text{false}\}$; \Comment{Target is not found.}
\end{algorithmic}
\end{algorithm}



\bibliographystyle{unsrt}
\bibliography{apssamp}

\end{document}